\documentclass[conference]{IEEEtran}
\IEEEoverridecommandlockouts
\usepackage{cite}
\usepackage{amsmath,amssymb,amsfonts}
\usepackage{algorithm}
\usepackage{algpseudocode}
\usepackage{graphicx}
\usepackage{textcomp}
\usepackage{array, booktabs, makecell}
\usepackage{xcolor}
\usepackage{caption}
\usepackage{subcaption}
\allowdisplaybreaks

\def\BibTeX{{\rm B\kern-.05em{\sc i\kern-.025em b}\kern-.08em
    T\kern-.1667em\lower.7ex\hbox{E}\kern-.125emX}}
\begin{document}
\title{Hardware-in-the-Loop Evaluation of Goodness of Fit (GoF) Testing for Dynamic Spectrum Sharing\\
\thanks{This work is supported through the UKRI/EPSRC Prosperity Partnership in Secure Wireless Agile  Networks (SWAN) EP/T005572/1.}
}

\author{\IEEEauthorblockN{Mir Lodro, Simon Armour, Mark A. Beach}
\IEEEauthorblockA{{Communication Systems and Networks Research Group}\\
\textit{School of Electrical, Electronic and Mechanical Engineering} \\
\textit{University of Bristol}\\
Bristol, United Kingdom \\
mir.lodro@bristol.ac.uk}
}

\maketitle

\begin{abstract}
In contrast to parametric spectrum sensing, non-parametric spectrum sensing can effectively detect the primary user's presence or absence without prior information about the primary user. Particularly, non-parametric spectrum sensing can be useful in dynamic spectrum sharing. The secondary user must detect incumbents and peer secondary users in dynamic spectrum sharing. The secondary user can use the licensed spectrum if the primary user is not detected using its band. The primary user detection problem is the goodness-of-fit testing problem. In this work, we performed a hardware-in-the-loop evaluation of goodness-of-fit tests such as Cramer-von-Mises (CM), Anderson-Darling (AD) and Kolmogorov-Smirnov (KS) tests. We used a wideband radio transceiver RFSoC 4x2 from AMD and an F8 radio channel emulator to perform GoF tests.
\end{abstract}

\begin{IEEEkeywords}
Goodness-of-fit tests, nonparametric sensing, software-defined radio, Keysight F8, goodness of fit, statistical sensing, RFSoC 4x2.
\end{IEEEkeywords}

\section{Introduction}
The need for high data rates continues to increase due to the rapid growth of low-cost sensors and B5G applications. Furthermore, 6G technology is expected to have a native spectrum sharing feature to make spectrum usage more efficient \cite{wang2023road}. This has led to growing interest in dynamic spectrum sharing for next-generation networks \cite{ghosh2023evolution}. Spectrum sharing is a solution to spectrum scarcity and underutilization, allowing secondary users to opportunistically share the primary user's spectrum. Efficient dynamic spectrum sharing requires robust spectrum sensing to identify the primary user (PU) band and avoid interference with the incumbents. In dynamic spectrum sharing, spectrum sensing is crucial for detecting the incumbents and other secondary users. The most common sensing method for primary user detection is energy detection. There is a rich source of background literature on spectrum sensing that focuses on energy detection (ED) \cite{gardner1991exploitation}. ED has the inherent limitation of poor performance in a low signal-to-noise ratio (SNR) \cite{tandra2008snr}. In addition to energy detection, other spectrum detection methods are matched filtering, cyclostationary spectrum detection \cite{han2006spectral}, and eigenvalue-based detection.
The matched filtering and cyclostationary sensing methods require information about the PU which is not always available in practice. Additionally, these spectrum-sensing methods are computationally inefficient. Cyclostationary sensing methods are prone to synchronization errors\cite{zeng2010robustness}.
The goodness-of-fit (GoF) test is a nonparametric sensing method to detect the presence or absence of a primary user in a band. In the GoF test, the spectrum decisions are made from noise sample distribution and signal-only distribution. The acquired samples are drawn from the noise sample distribution without a primary user. The work in \cite{teguig2015spectrum} presents spectrum sensing and likelihood goodness of fit test. We have organized our work into six sections. After the introduction in Section I, we explain the spectrum-sensing scenario in Section II. Section III briefly explains the Goodness of Fit test. Section IV discusses the hardware setup used for the emulation and the data acquisition for spectrum sensing. The results and discussion are explained in Section V. In Section VI, we conclude our work.

\section{Spectrum Sensing Scenario}
In a dynamic spectrum sensing scenario, the secondary user (SU) must detect not only the presence of primary user devices but also the transmission of another secondary user network. For example, in a hybrid spectrum sharing scenario, to avoid interference with the International Mobile Telecommunications (IMT) device, the WiFi access point (AP) must detect the presence of IMT cellular before it starts transmission to the WiFi station (STA). Another example is the coexistence of the tactical data link (TDL) and electronic news gathering (ENG) in \(2025-2110\) MHz \cite{mchenry2024hybrid}. It can also complement the database approach of spectrum sharing.
\begin{figure}
    \centering
    \includegraphics[width=\columnwidth]{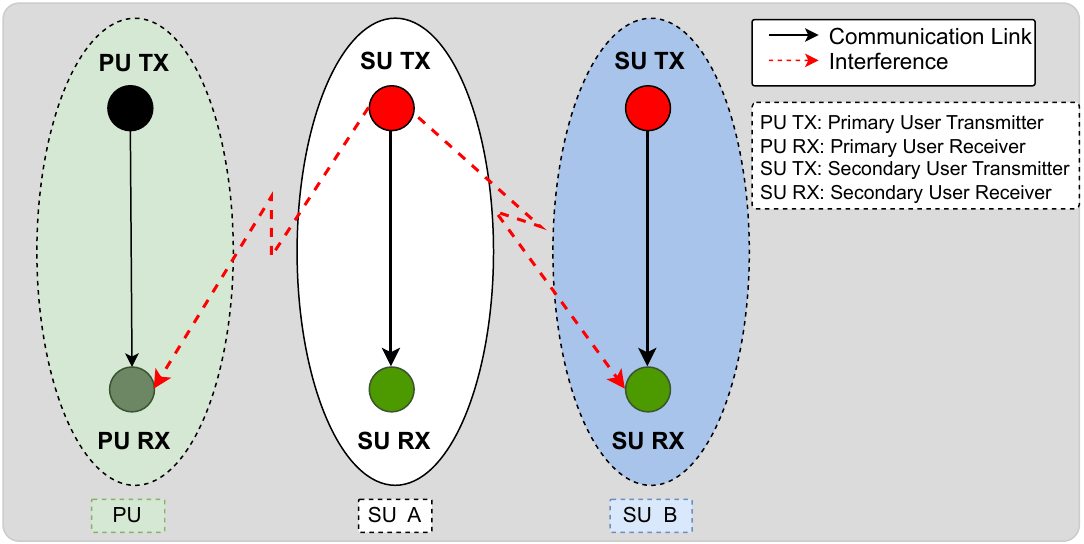}
    \caption{Schematics of spectrum sensing and the SU coexistence with incumbents and peer SU network.}
    \label{fig:sharing}
\end{figure}
Fig. \ref{fig:sharing} shows spectrum sensing in a dynamic spectrum sharing scenario. If the link between the SU and the primary user transmitter and another secondary user transmitter of network B is poor, that is, deeply shadowed, the SU transmitter (TX) will not detect it. Upon missed detection of the PU link and another secondary user link, the SU TX of the secondary user network A will perform transmission to its receiver. The transmission of SU Tx A will create interference to the primary user network as well as to the receiver of secondary user network B. Hence, spectrum sensing is critical for a more dynamic spectrum sharing.
\section{Goodness-of-Fit (GoF) Tests}
The GoF tests are nonparametric methods to know the presence of the primary user using its allocated channel. The procedure for detecting the PU signal involves the computation of the empirical cumulative distribution function (CDF) \(F_1(y)\) of the test statistics of the acquired samples and the CDF \(F_0(y)\) of the noise-only samples.  PU is detected if the difference between the reference CDF \( F_0(y)\) and the empirical CDF \( F_1(y) \) exceeds a threshold. This also satisfies the Kullback-Leibler (KL) divergence test which quantifies the closeness of the two density functions. The difference between two densities is 0 if the two densities match exactly.

\subsection{Kolmogorov-Smirnov}
The Kolmogorov-Smirnov (KS) test is a nonparametric method to determine whether the acquired samples are drawn from the noise distribution \(F_0 (y)\). The test statistic is developed by measuring the maximum absolute difference between the CDFs of the test statistics of the data samples and the test statistic of the noise samples \cite{zhang2010fast}. For the finite sample size \(N\) the maximum difference between the two CDFs is given as:
\begin{equation}
D_{Y}=\sup \left\{\left|F_{1}(y)-F_{0}(y)\right|:0<y<N-1\right\}
\end{equation}
The null hypothesis \( \mathcal{H}_{0}\) is only rejected if the test statistic exceeds the threshold that satisfies the probability of false alarm \( P_{fa} \).

\subsection{Cramer-von-Mises and Anderson-Darling}
The Cramer-von Mises GoF test is investigated in \cite{kieu2011cramer}. The Cramer-von Mises test statistic \(W^2\) is shown in eq.(\ref{eq:cm}):
\begin{equation}
    W^{2} \triangleq n \int_{-\infty}^{+\infty}\left(F_{1}(y)-F_{0}(y)\right)^{2} d F_{0}(y)
    \label{eq:cm}
\end{equation}
It can be seen that eq. (\ref{eq:cm}) does not give enough weight to the tails of the distribution. Anderson and Darling introduced a more flexible solution and introduced the weight function \(\phi(F_0(y))\) to the CM test statistic. Hence, the modified AD GoF test statistic \(A_{c}^{2}\) is given in eq.(\ref{eq:ad}).

\begin{equation}
        A_{c}^{2} \triangleq n \int_{-\infty}^{+\infty}\left(F_{1}(y)-F_{0}(y)\right)^{2} \phi(F_{0}(y)) d F_{0}(y)
    \label{eq:ad}
\end{equation}
The probability of false alarm \(P_{fa}\) using AD test when null-hypothesis is true \(\mathcal{H}_0\) is given as: 
\begin{equation}
    P_{fa}=Pr\{A_{c}^{2}>t_0|H_0\}=\alpha
\end{equation}
Where $\alpha$ is the desired false alarm probability.
The probability of detection \(P_{d}\) using AD test for the critical value $t_0$ when alternative hypothesis \(\mathcal{H}_1\) is true is given as:
\begin{equation}
    P_{d}=Pr.\{A_{c}^{2}>t_0|H_1\}=1-F_{{Ac}^{2}}(t_0)
\end{equation}

\section{Hardware Setup}
The measurement setup used is shown in Fig. \ref{fig:hardware}. The hardware setup consists of a Zynq UltraScale+ Gen3 RFSoC 4x2-a wideband direct conversion transceiver from AMD and a Keysight F8 radio channel emulator. The RFSoC 4x2 has two transmit and four receive channels. The RFSoC 4x2 can transfer samples in the range of 16 to 32768. The RFSoC 4x2 has a processing system (PS) and programmable logic (PL) for high-speed signal processing. The PS runs the PYNQ framework with Jupyter Notebook which controls the interaction between the PS and the PL. The baseband PU signal is generated using the RFSoC 4x2 processing system. The primary user signal is a continuous wave (CW) tone with a baseband frequency of 100 kHz. The sampled PU signal is upconverted, amplified, and prepared to be fed into port 1 of the radio channel emulator. DAC B port of the RFSoC 4x2 was used for signal transmission. DAC A of the RFSoC 4x2 was terminated with a \( 50 \,\Omega \) load. Four output ports of the channel emulator are connected to four Rx ports of RFSoC 4x2. Each channel of RFSoC 4x2 is considered a secondary user. The RFSoC 4x2 is configured to operate in the receive mode and acquire complex samples of size 4096. The ADCs in each channel are configured to operate at a sample rate of 4.9152 GSPS and the interpolation and decimation factors were set to 2. Hence, the sampling rate for each channel of RFSoC 4x2 is 2.4576 GSPS. The step attenuation of ADC D can be controlled from 0 to 27, where 0 means the lowest attenuation and 27 the highest attenuation. The step attenuation of the ADC block is set to 0 and there is no offset between the in-phase (I) and quadrature (Q) samples of the acquired samples. The samples obtained are rearranged to form a sample covariance matrix. The eigenvalues are measured from the covariance matrix of the computed sample. The test statistics are developed for different variants of eigenvalue detection \cite{zeng2007maximum}. For maximum eigenvalue (\(\lambda_{\mathrm{max}}\)), minimum eigenvalue (\(\lambda_{\mathrm{min}}\)) and \(i\mathrm{th}\) eigenvalue \((\lambda_\mathrm{i}\)), the test statistics (\(\Gamma_{}\)) of the variants of the detection of eigenvalues can be given as:
\begin{figure}
    \centering
    \includegraphics[width=0.98\columnwidth]{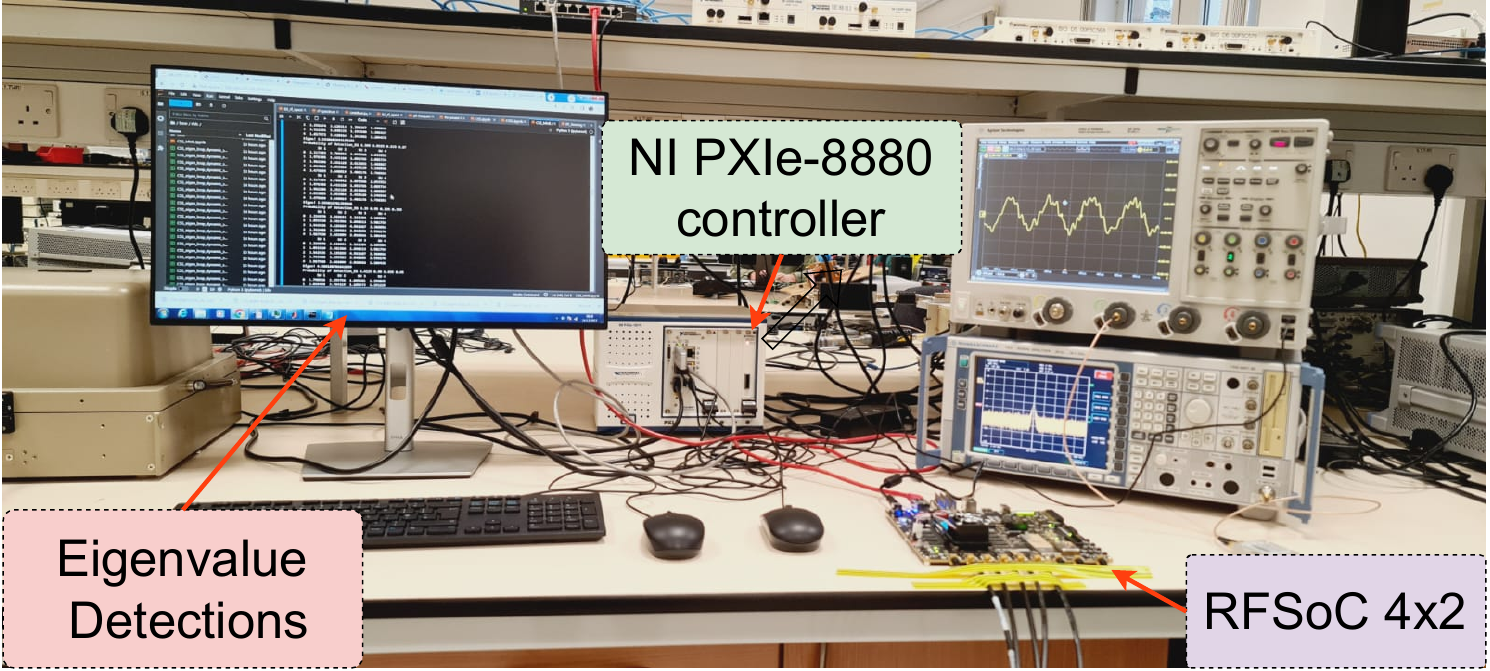}
    \caption{Measurement setup.}
    \label{fig:hardware}
\end{figure}
\begin{itemize}
    \item Maximum Minimum Eigenvalue (MME): \begin{equation}
    \Gamma_{MME}=\frac{\lambda_{\text{max}}}{\lambda_{\text{min}}}
    \end{equation}
    \item Maximum Eigenvalue to Arithmetic Mean (ME-AM): \\
    \begin{equation}
       \Gamma_{\mathrm{ME-AM}}= \frac{\lambda_{\text{max}}}{\frac{1}{N}\sum_{i=1}^{N} \lambda_i}
    \end{equation}
    \item Maximum Eigenvalue to Geometric Mean (ME-GM): 
    \begin{equation}
    \Gamma_{\mathrm{ME-GM}}=\frac{\lambda_{\text{max}}}{\left(\prod_{i=1}^{N} \lambda_i\right)^{\frac{1}{N}}}    
    \end{equation}
    
    \item Arithmetic Mean to Geometric Mean (AM-GM): 
    \begin{equation}
    \Gamma_{\mathrm{AM-GM}}    \frac{\frac{1}{N}\sum_{i=1}^{N} \lambda_i}{\left(\prod_{i=1}^{N} \lambda_i\right)^{\frac{1}{N}}}
    \end{equation}

\end{itemize}
The samples received at each SU are subject to channel profile replayed by the radio channel emulator. The measurements are conducted at the center frequency of \( 1800\, \mathrm{MHz}\) and varying PU transmit power levels. The transmit power of the PU is controlled by changing the DAC variable output power (VOP) values of the RFSoC 4x2.

\section{Results and Discussions}
\begin{figure}[H]
    \centering
    \includegraphics[width=\columnwidth]{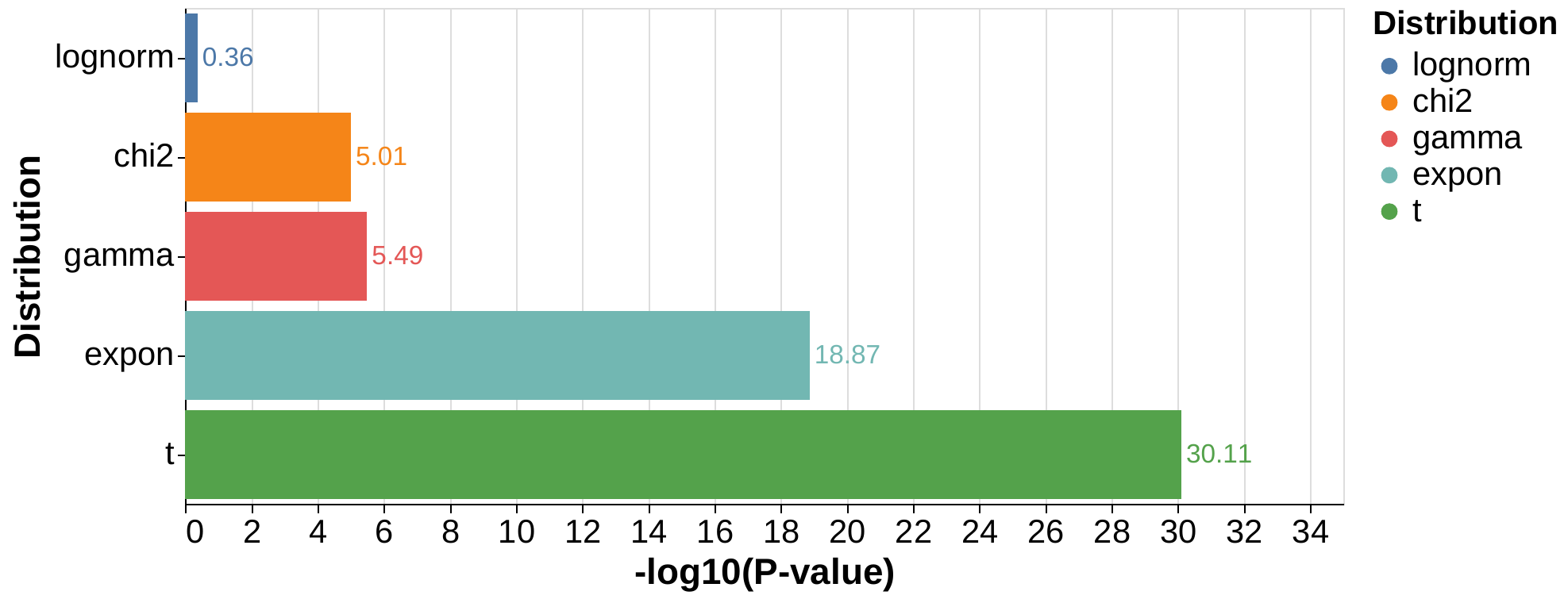}
    \caption{P-values of the candidate distributions.}
    \label{fig:p-value}
\end{figure}

\begin{figure}[H]
    \centering
    \includegraphics[width=\columnwidth]{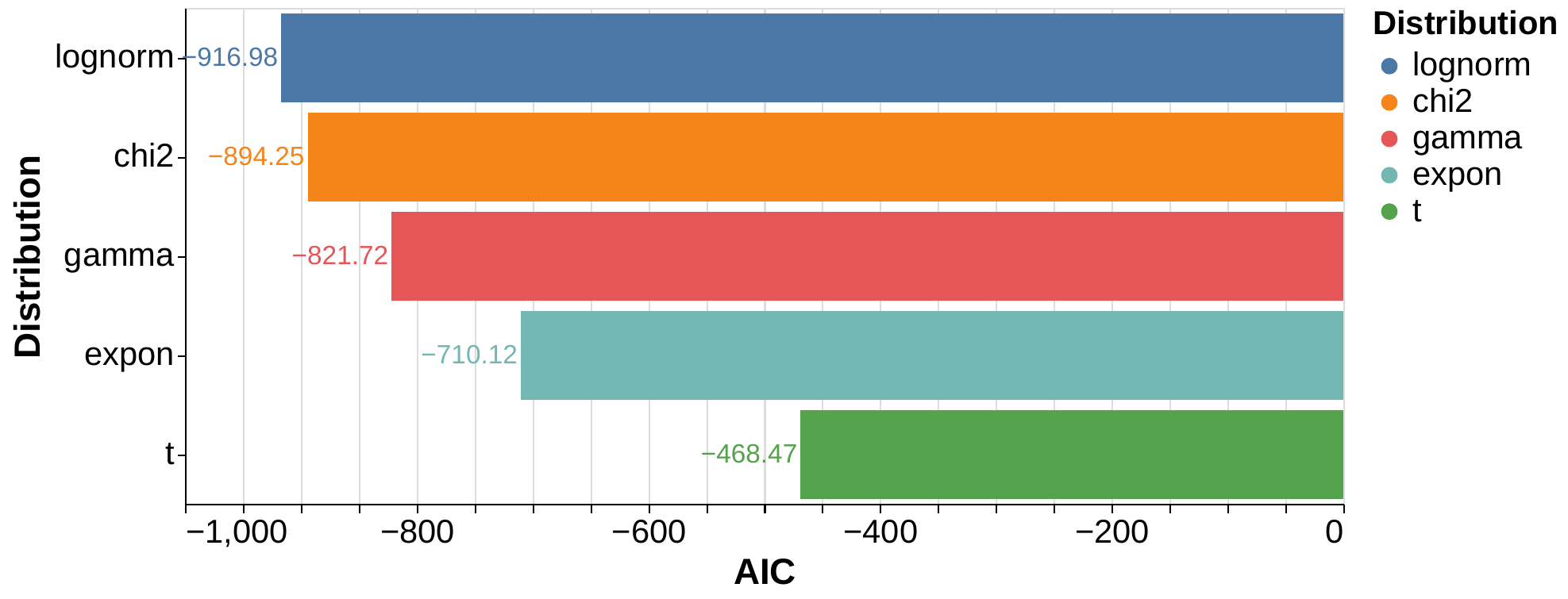}
    \caption{AIC values of the candidate distributions.}
    \label{fig:aic-value}
\end{figure}

In this section, we present the test statistics of noise samples i.e. when the PU is absent, and the test statistics of the data samples i.e. when the PU is present. We present the test statistics for noise-only and data samples using four variants of eigenvalue detection.
\begin{figure}
    \centering
    \includegraphics[width=\columnwidth]{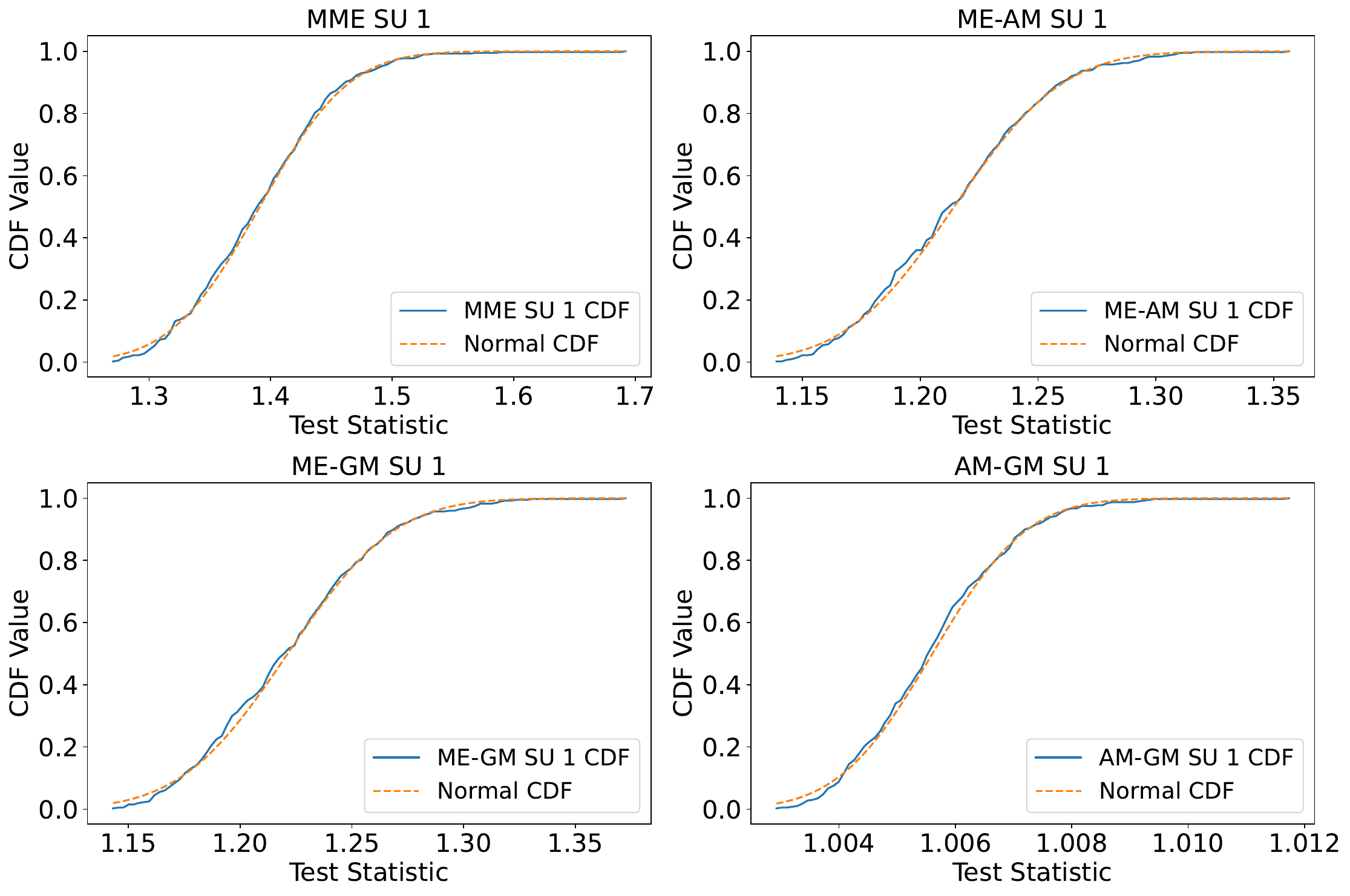}
    \caption{ Distribution fitting and the CDF plots of the noise data at SU 1. The test statistics of noise data are measured using MME, ME-AM, ME-GM, and AM-GM.}
    \label{fig:noise_CDFs}
\end{figure}
Fig. \ref{fig:noise_CDFs} shows the CDFs of the noise data test statistics using four eigenvalue sensing methods. The empirical distribution of the test statistics obtained from the measured noise samples follows the theoretical noise distribution. It also satisfied the Kullarback-Leibler (KL) divergence test where the difference between two densities $P$ and $Q$ related by the following expression tends to zero when the two densities match exactly.
\begin{equation}
D_{\text{KL}}(P \parallel Q) = \sum_{i} P(i) \log \left( \frac{P(i)}{Q(i)} \right) \approx 0
\end{equation}
Hence, the null hypothesis can not be rejected. In the second set of measurements, we recorded samples at four channels of RFSoC 4x2 in the presence of an RF signal. The RF signal was generated using channel 1 of RFSoC 4x2 and fed to the F8 radio channel emulator input port 1. The RF channel emulator is configured to playback the 7-tap extended pedestrian A (EPA) channel model. We can see when the spectrum is occupied the KL test is not satisfied i.e. $D_{\text{KL}}(P \parallel Q) > 0$ which means that the alternative hypothesis \( \mathcal{H}_1\) is true. 
\begin{figure}
    \centering
    \includegraphics[width=\columnwidth]{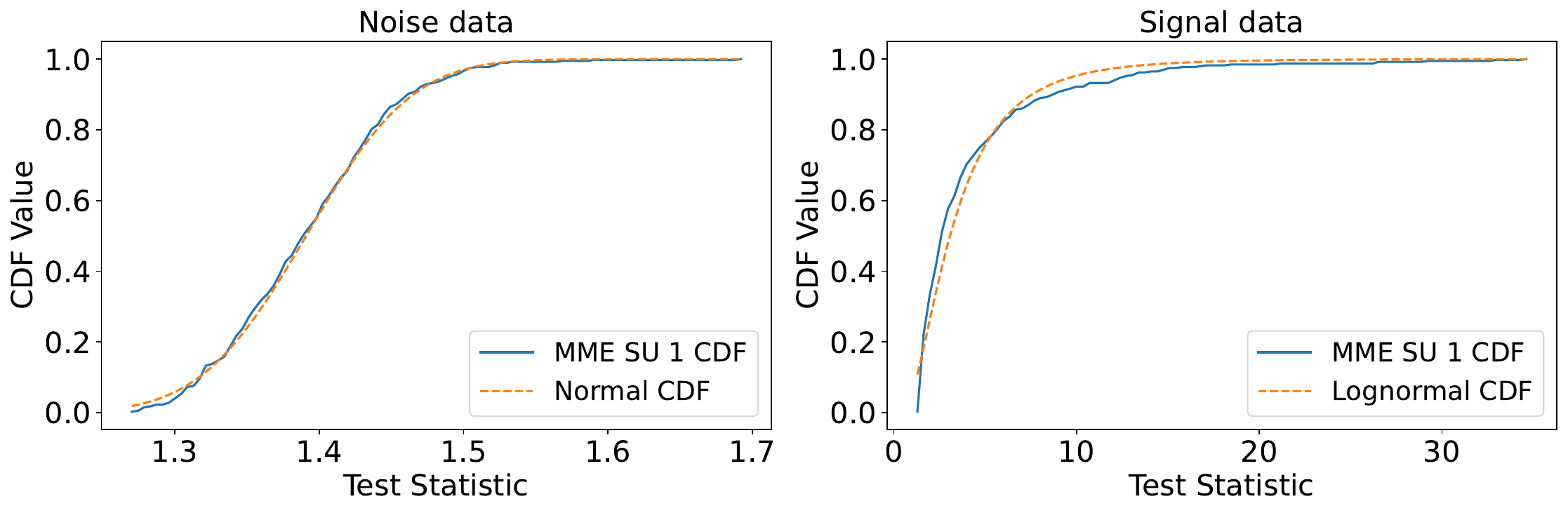}
    \caption{CDF plots and distribution fitting of MME test statistics of the noise samples and signal samples.}
    \label{fig:signal_CDFs}
\end{figure}
Fig. \ref{fig:signal_CDFs} shows the noise and signal data CDFs in SU 1 using MME eigenvalue detection. We can see that the CDFs of the test statistics of the eigenvalue of the signal data do not follow the Gaussian distribution. Furthermore, \( F_0(y) < F_1(y)\), which confirms the presence of PU. It shows the empirical distributions using eigenvalue detection schemes do not follow noise distribution. Hence, the null hypothesis $\mathcal{H}_0$ can be rejected in favor of the alternative hypothesis $\mathcal{H}_1$. Thus, we validate the true case that the primary user is present.
\begin{figure}
    \centering
    \includegraphics[width=\columnwidth]{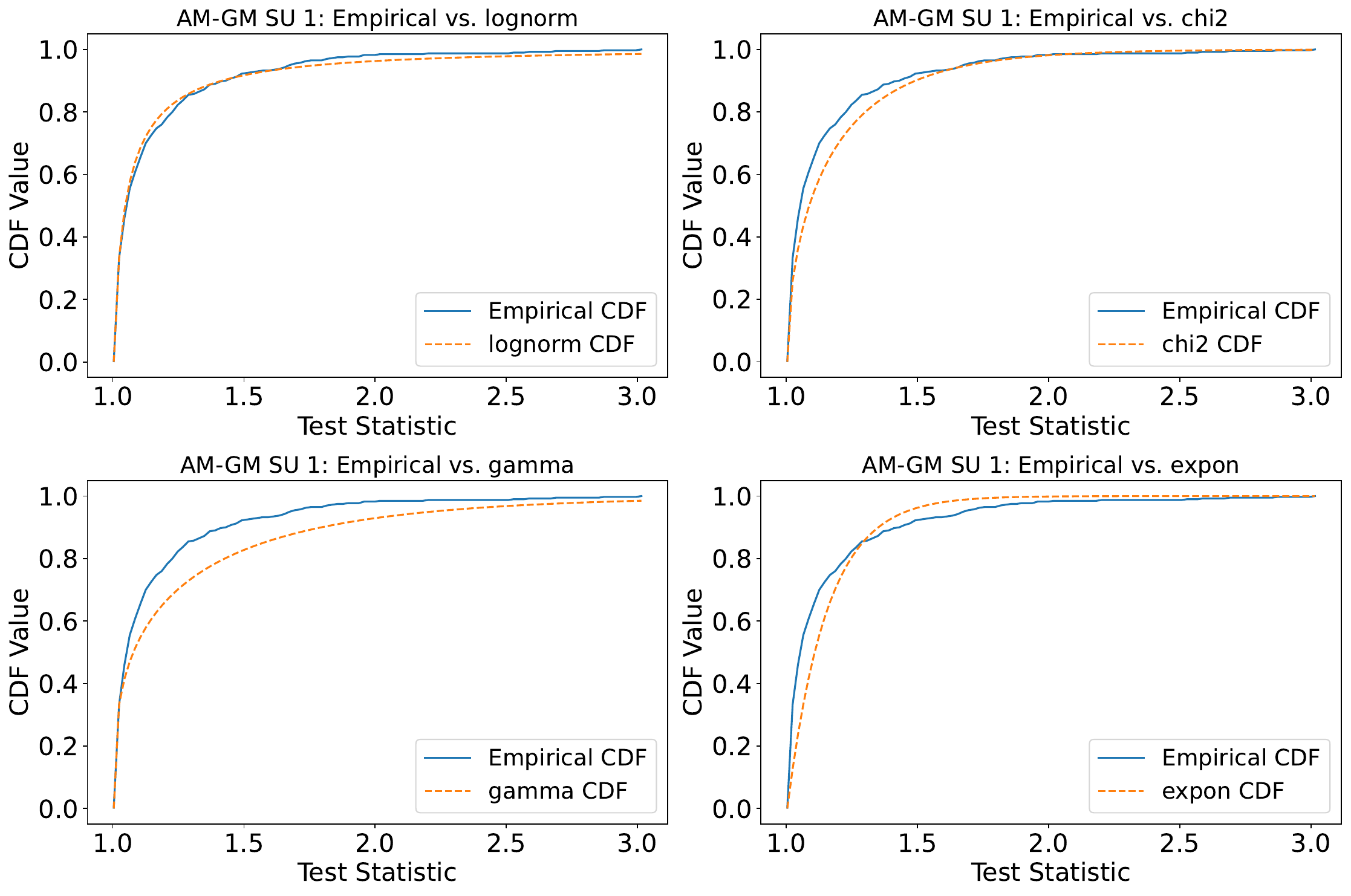}
    \caption{CDF plot and distribution fitting of the signal data of SU 1 using AM-GM eigenvalue detection. Candidate distributions applied are lognormal, Chi-square, gamma, and exponential distribution.}
    \label{fig:CDF_fitting}
\end{figure}
We also performed a distribution fitting of the data with candidate distributions. The selected candidate distributions are lognormal, Chi-Square, Gamma, and Exponential. Akaike Information Criterion (AIC) is performed for each fitted distribution. We calculate the AIC using the number of parameters estimated $k$ and $L$ the maximum value of the log-likelihood function. 
\begin{equation}
    AIC=2k-2ln(L)
\end{equation}
AIC is a goodness of fit testing to measure the accuracy of fitting the statistical model with the data. We choose the distribution with the lowest AIC. Fig. \ref{fig:CDF_fitting} shows the CDF plots of the SU 1 signal data using four eigenvalue detection methods. Each subplot in Fig. \ref{fig:CDF_fitting} compares the empirical CDF and the theoretical CDF of the best-fit distribution. Based on the AIC and visual inspections, the best-fitting distribution is the lognormal distribution as the CDF closely matched with the empirical CDF.
The P value and AIC value for different distributions have been quantified in Fig. \ref{fig:p-value} and Fig. \ref{fig:aic-value} respectively.
It can be seen from the bar charts in Fig. \ref{fig:p-value} that the lognormal distribution performs best with a P-value of $0.36$ followed by a Chi-Square distribution with a P-value of $5.01$. In addition to the P-value, the fitting is also confirmed with the lowest AIC values for the lognormal distribution. It can be seen from Fig. \ref{fig:aic-value} that the AIC values also confirm that the best fit with the data samples is the log-normal distribution with the lowest AIC value of $-916.98$. 

\begin{table}[htbp]
    \centering
    \caption{AD Test Statistics and Critical Values}
    \label{tab:order_critical_values}
    \begin{tabular}{l|cccc}
        \toprule
        \textbf{Parameters}& \textbf{SU 1}  \\
        \midrule
        AD Test Statistic & 46.89   \\
        Critical Value at 15\% &0.57  \\
        Critical Value at 10\% &0.65  \\
        Critical Value at 5\% &0.78  \\
        Critical Value at 2.5\% &0.90  \\
        Critical Value at 1\% &1.08 \\
        \bottomrule
    \end{tabular}
\end{table}
The test statistic for each SU is significantly higher than the critical values \(0.57,0.65,0.78, 0.90 \,\mathrm{and}\, 1.08\) measured at significance levels of \(15\%,10\%,5\%, 2.5\%, \mathrm{and}\, 1\%\) respectively. It can be seen that the data samples don't follow normal distribution at these significance levels.
\section{Conclusion}
Hardware emulation of nonparametric spectrum sensing is performed using an RF channel emulator to detect the presence or absence of the primary user. We used the RF channel emulator and RFSoC 4x2 with high-speed ADCs to capture the data samples for both cases where there was no primary user signal and when the primary user signal was present. The RF channel emulator was configured to emulate the 3GPP EPA channel model between PU and all SUs. The test statistics for different variants of eigenvalue detection algorithms are used for the performance evaluation of GoF nonparametric spectrum sensing. We validated the nonparametric spectrum sensing test perfectly fitting the noise-only case and validating the presence of weak PU signal from the data samples.
\section*{Acknowledgment}

This work is supported through the UKRI/EPSRC Prosperity Partnership in Secure Wireless Agile Networks (SWAN) EP/T005572/1.


\bibliographystyle{IEEEtran}

\bibliography{main_GoF.bib} 


\end{document}